\documentstyle[amsfonts,12pt]{article}

\begin{document}

\title{Long-time asymptotic of temporal-spatial coherence function for light
propagation through time dependent disorder}
\author{M. Auslender \\
Department of Electrical and Computer Engineering\\
Ben-Gurion University of the Negev\\
P.O.B. 653, Beer-Sheva, 84105 Israel\\
e-mail: marka@newton.bgu.ac.il}
\date{}
\maketitle
\thispagestyle{empty}
\begin{abstract}
Long-time asymptotic of field-field correlator for radiation propagated 
through a medium composed of random point-like scatterers is studied using 
Bete-Salpeter equation. It is shown that for plane source the fluctuation 
intensity (zero spatial moment of the correlator) obeys a power-logarithmic 
stretched exponential decay law, the exponent and preexponent being dependent 
on the scattering angle. Spatial center of gravity and dispersion of the 
correlator (normalized first and second spatial moments, respectively) prove 
to weakly diverge as time tends to infinity. A spin analogy of this problem is 
discussed.

\end{abstract}

\newpage
\baselineskip 24pt

\begin{center}
{\Large {\bf 1. Introduction}}
\end{center}

The fluctuation phenomena in scattering of electromagnetic waves propagating 
in random media was the subject of extensive studies during the last decade. 
This activity was inspired by experimental evidence for a number of the phenomena
(e.g. enhanced back-scattering [1,2], spatial and frequency correlation of
back-scattered and transmitted radiation [3,4]; for a review see [5]) as well
as by realizing a close similarity of these phenomena to fluctuation
phenomena in electron transport in disordered systems (mesoscopics) [6,7].
The latter brought new physical insight and theoretical methods into the
field established in optics many years ago [8]. 
The study of fluctuation effects was extended also to dynamic disorder in
which the scatterers move at random. Golubentsev showed that the
contribution of interference back-scattering to static field auto-correlator
(in particular, average intensity), which is essential for immobile
scatterers, is suppressed when they move [9]. The similar result for
temporal-spatial auto-correlator (dynamic coherence function) was obtained
by Stephen [10]. Short-time fluctuations were studied using diffusion 
approximation in [10] for the cases of a point source in an infinite medium 
and a plane wave incident on semi-infinite medium (Milne problem). However,
the validity of this approximation for time-dependent disorder remained
questionable and deserves special study, to which the present paper is
devoted.

\begin{center}
{\Large {\bf 2. Definitions and Statement of the Problem}}
\end{center}

Let the electric field $ E\left({\bf r},t\right) $ of light obeys scalar wave 
equation with a source current density $j({\bf r},t)$. Coherent field (CF) and 
field coherence function (FCF) are defined respectively by 
\begin{equation}
E_c\left( {\bf r},t\right) =\langle E\left( {\bf r},t\right) \rangle ,
\end{equation}
\begin{equation}
\Gamma \left({\bf r}_1,t_1;{\bf r}_2,t_2\right) =
\langle E\left({\bf r}_{1},t_{1}\right)E^{\ast}\left({\bf r}_2,t_2\right)\rangle 
-\langle E\left({\bf r}_{1},t_{1}\right)\rangle 
\langle E^{\ast}\left({\bf r}_2,t_2\right)\rangle ,  
\end{equation}
where brackets mean average over realizations of underlying random process.
In the case under consideration it is the scatterers motion that causes the
randomness. Point-like scatterers are considered, so individual scattering 
can be described by scattering amplitude $f_{s}$ (see e.g. [8], vol.1). In 
weak multiple scattering limit $\lambda \ll l$ (where $\lambda $ is the 
radiation wavelength, $l=4\pi /n_s|f_s|^2$ is radiation mean free path, and 
$n_s$ is the scatterers density) CF is easily calculated. For space-time 
Fourier transform of CF one obtains 
\begin{equation}
E_c\left( {\bf p},\omega \right) =\frac{4\pi ik_0}{c}
j\left( {\bf p},\omega\right) G\left( {\bf p},\omega \right) ,\quad G\left( {\bf p},\omega \right)
=\frac{1}{{\bf p}^2-k_0^2-n_sf_s}\,,\
\end{equation}
where $ k_{0}=2\pi /\lambda $ is the wave number, $\omega =ck_{0} $ is the
radiation frequency and $ G({\bf p},\omega) $ is `one-particle` Green
function (see e.g. [6,7]). When considering FCF slow coordinate 
${\bf r}=({\bf r}_1+{\bf r}_2)/2$ and time $t=(t_1+t_2)/2 $ as well as fast 
ones: ${\bf \rho }=({\bf r}_1-{\bf r}_2)/2$ and $\tau =(t_1-t_2)/2$ are 
introduced. Then Fourier transform of FCF (Eq.2) with respect to the fast
variables 
\begin{equation}
\hat{\Gamma }\left( {\bf r},t;{\bf p},\omega \right) = 
\int\!\int_{-\infty}^{\infty}\exp \left( i\omega \tau -i{\bf p}\cdot {\bf %
\rho }\right) \Gamma \left( {\bf r}_1,t_1;{\bf r}_2,t_2\right) \frac{d^3\rho
 d\tau }{\left( 2\pi \right) ^4}  
\end{equation}
is commonly used. This quantity may be interpreted as energy density in
incoherent portion of the radiation with the frequency $\omega $ and 
propagation direction ${\bf n}={\bf p}/p$ at the point ${\bf r}$ and 
time instant $t$ (see e.g. [8], Vol. 2).

Main issue of the present paper is the study of $\hat{\Gamma}\left({\bf r},t;{\bf p},\omega \right)$ 
at long times neglecting the coherence effects. By 'long times' I mean the regime when 
$\Delta \left( t\right)$, mean square displacement a scatterer moves during time $t$, is lagrer than 
the wavelength: 
\begin{equation}
\Delta \left( t\right) >\lambda.
\end{equation}
 
\begin{center}
{\Large {\bf 3. Bete-Salpeter Equation}}
\end{center}

Bete-Salpeter equation in weak individual scattering regime was derived in
[10]. In general case, except some parametric factors, the structure of this 
equation remains unchanged. For $\hat{\Gamma}\left( {\bf q},t;{\bf p}\right)$, 
spatial Fourier transform of $\hat{\Gamma}\left({\bf r},t;{\bf p}\right) $, 
it reads \ 
\[
\hat{\Gamma}\left( {\bf q},t;{\bf p}\right) =n_s|f_s|^2G\left( {\bf p}-{\bf q%
}/2\right) G^{\ast }\left( {\bf p}-{\bf q}/2\right) \times 
\]
\begin{equation}
\int \exp \left[ -\frac 12\Delta \left( t\right) ^2|{\bf p}-{\bf p}%
_1|^2\right] \left[ \hat{\Gamma}\left( {\bf q},t;{\bf p}_1\right) -E_c\left( 
{\bf p}_1-{\bf q}/2\right) E_c^{\ast }\left( {\bf p}_1+{\bf q}/2\right)
\right] \frac{d^3p_1}{\left( 2\pi \right) ^3}\,.  
\end{equation}
The argument $\omega $ here and further is omitted for brevity. It is
worth noting that Eq.6 may be also derived using Foldy-Twersky ansatz (see
[8], Vol. 2). At distances much larger than wavelength, that is at $q/k_0\ll 1$, 
the approximation
\begin{equation}
G\left( {\bf p}-{\bf q}/2\right) G^{\ast }\left( {\bf p}-{\bf q}/2\right)
\simeq \frac{\pi l}{2k_0^2}\cdot \frac{\delta (|{\bf p}|-k_0)}{1-il{\bf q}
\cdot{\bf p}/k_0}  
\end{equation}
holds well, and Eq.6 with account of Eq.3 will be satisfied by the
substitution 
\begin{equation}
\hat{\Gamma}\left({\bf q},t;{\bf p}\right) =\frac{(2\pi)^3\,l}{c^2}
\delta\left( |{\bf p}|-k_0\right)\gamma\left({\bf q},t;{\bf p}/k_0\right) 
\end{equation}
that transforms Eq.6 to the transport equation for the {\em on-shell} FCF 
$\gamma \left( {\bf q},t;{\bf n}\right)$ 
\[
\left(1-il{\bf q}\cdot {\bf n}\right) \gamma \left( {\bf q},t;{\bf n}\right) 
-\int_{\left| {\bf n}_{1}\right|=1}\exp \left[-\frac{{\cal J}}{2}|{\bf n}-{\bf n}_1|^2\right] 
\gamma\left( {\bf q},t;{\bf n}_1\right)\frac{d\Omega _{1}}{4\pi} \]
\begin{equation}
=\int_{|{\bf n}_0|=1}\exp\left[-\frac{{\cal J}}2|{\bf n}-{\bf n}_0|^2\right] 
\frac{j\left( k_0{\bf n}_0-{\bf q}/2\right) j^{\ast}\left(k_0{\bf n}_0+{\bf q}/2\right)}
{1-il{\bf q}\cdot {\bf n}_0}\cdot\frac{d\Omega _0}{4\pi}
\,,\quad {\cal J}=\left( k_0\Delta\right)^2.\,  
\end{equation}
\\For point source Eq.9 coincides with that derived by Golubentsev [9] using a 
special diagram technique whose rules take explicitly into account Eq.7. The sources of 
other types were  considered [11] within real-space approach (RSA) after Rosenbluh et al.
[12]. In the next section a connection between Bete-Salpeter equation and
RSA will be discussed.

\begin{center}
{\Large {\bf 4. Real-Space Trajectory Decomposition \\and Spin Analogy}}
\end{center}

Solving Eq.9 by iterations of the integral term represents its solution as
an average of $\gamma \left({\bf q},t;{\bf n}_0,{\bf n}\right)$, source-detector 
directional FCF, over the propagation directions ${\bf n}_0$ allowed for radiation
\begin{equation}
\gamma ({\bf q},t;{\bf n})=\int\limits_{|{\bf n}_0|=1}j\left( k_0{\bf n}_0-%
{\bf q}/2\right) \,j^{\ast }\left( k_0{\bf n}_0+{\bf q}/2\right) \gamma
\left( {\bf q},t;{\bf n}_0,{\bf n}\right) \frac{d\Omega _0}{4\pi }  
\end{equation}
with the series expansion in the integrand: 
\[
\gamma ({\bf q},t;{\bf n}_0,{\bf n})=
\]
\begin{equation}
\sum_{N=2}^\infty \underbrace{\int
\ldots \int }_{N-1}W_N\left( {\bf q};{\bf n}_0,{\bf n}_1,\ldots ,{\bf n}_N\right)%
\exp \left[-\frac{{\cal J}}2\sum_{j=1}^N|{\bf n}_{j-1}-{\bf n}_j|^2
\right]\frac{d\Omega^{N-1}}{\left(4\pi\right)^{N-1}}\,,\;{\bf n}_N={\bf n\,.}  
\end{equation}
Here $W_N$ are scattering configuration weights given by 
\begin{equation}
W_N\left( {\bf q};{\bf n}_0,{\bf n}_1,\ldots ,{\bf n}_N\right)
=\prod_{j=0}^N\frac 1{1-il{\bf q}\cdot {\bf n}_j}\,.  
\end{equation}
At $N\gg 1$ and $|{\bf r}|\gg l$ the inverse transform Fourier of this
product is strongly peaked at ${\bf r}=l\sum_{j=0}^N{\bf n}_j$, so in real 
space FCF appears to be a sum of path integrals over polygon trajectories
terminated at the point ${\bf r}$. Each $j$th ($0\leq j\leq N$)
straight-line part of them has the length $l$ and directed along the vector 
${\bf n}_j$. The initial ${\bf n}_0$ and final ${\bf n}_N$ directions are
determined by a source and a detector, respectively, and each change of 
'free-flight' direction occurs after a scattering act. Scattered directions 
${\bf n}_1,\ldots ,{\bf n}_{N-1}$ are distributed with correlation if 
${\cal J}$ $\neq 0$. 

Thus, basic assumptions of RSA [12] result from the Bete-Salpeter equation. 
In RSA, however, $W_N$ are calculated using ${\em apriori}$ random-walk 
considerations, and the exponential weights due to scatterers motion are 
approximately averaged by averaging of their exponents [12]. An analysis 
shows that $W_N$ of RSA is an average of right-hand side of Eq.12 at 
$ql\ll 1$ and $N\gg 1$ over fully random ${\bf n}_j$. With the only this less 
stringent approximation $\gamma \left( {\bf q},t;{\bf n}_0,{\bf n}\right)$ 
will coincide with conditional partition function in an ensemble of one-dimensional 
classical Heisenberg spin systems, each with the same exchange integral ${\cal J}$,
but different numbers $N\geq 2$ of species and weights $W_N$. This analogy
facilitated an estimation of FCF within RSA making use the methods of
magnetism [13].

\begin{center}
{\Large {\bf 5. Long-Time Asymptotic of Coherence Function}}
\end{center}

Scatterers motion removes diffusion singularity: $\gamma ({\bf q},t;{\bf n}%
_0,{\bf n})\propto q^{-2}$ at ${\bf q}\rightarrow 0$ specific for immobile
scatterers. At $ql\ll 1$ it makes sense to consider hydrodynamic expansion 
\begin{equation}
\gamma \left( {\bf q},t;{\bf n}_0,{\bf n}\right) =g_{0}({\bf n}_0,{\bf n}%
)+\sum_{k=1}^{\infty}i^k\sum_{{\bf \alpha }}\frac{q_{\alpha _1}...q_{\alpha _k}}{%
k!} g_{\alpha _1...\alpha _k}({\bf n}_0,{\bf n}),  
\end{equation}
where $\alpha $'s are the Cartesian indices, and the functions $g_{\alpha
_1...\alpha _k}\left( t;{\bf n}_0,{\bf n}\right) $ 
\begin{equation}
g_{\alpha _1...\alpha _k}\left( {\bf n}_0,{\bf n}\right) =\left. i^{-k}\frac{%
\partial ^k\gamma \left( {\bf q},t;{\bf n}_0,{\bf n}\right) }{\partial
q_{\alpha _1}...\partial q_{\alpha _1}}\right| _{{\bf q}=\,0}=\int r_{\alpha
_1}...r_{\alpha _k}\gamma \left( {\bf r},t;{\bf n}_0,{\bf n}\right) d^3r 
\end{equation}
are spatial moments. The first term $g_{0}\left( {\bf n}_0,{\bf n}\right)$ 
of the expansion in Eq.13 (the zero moment) is the intensity of the 
fluctuations in a macroscopic volume. Next two moments normalized to the zero one 
have clear physical meanings. The vector and matrix 
\begin{equation}
\overline{r_\alpha }=\frac{g_{\alpha} \left( {\bf n}_0,{\bf n}\right) }{%
g_{0}\left( {\bf n}_0,{\bf n}\right) },\quad \overline{r_\alpha r_\beta }=%
\frac{g_{\alpha \beta }\left( {\bf n}_0,{\bf n}\right) }{g_{0}\left( {\bf n}_0,%
{\bf n}\right) }  
\end{equation}
are the spatial center of gravity and dispersion matrix, respectively. 
Eq.13 can be directly obtained by iterations of the term 
$\propto {\bf q}\cdot {\bf n}$ in the transport equation for 
$\gamma ({\bf q},t;{\bf n}_0,{\bf n})$, which differ from Eq.9 only in the 
right-hand side. Such a procedure creates recurrence equations for the 
moments, which enable us to calculate them recursively, but analyzing the full 
set of the moments (same as solving the original equation) is hardly possible. 
Nevertheless, the asymptotics of the above mentioned three
moments at ${\cal J}\rightarrow \infty $ can be calculated rigorously. This limit 
is quite reasonable since Eq.5 implies the strong inequality ${\cal J}\gg 1$.
\begin{center}
{\large {\bf 5.1. Fluctuations Intensity}}
\end{center}

Using Eq.10 and Eq.9 one obtains the equation for the zero moment 
\begin{equation}
\left( 1-{\frak L}\right) g_{0}\left( {\bf n}_0,{\bf n}\right) =\exp \left[ -%
\frac{{\cal J}}2|{\bf n}-{\bf n}_0|^2\right] ,  
\end{equation}
where ${\frak L}$ is the same integral operator as appears in the left-hand side
of Eq.9. The eigenfunctions of the operator ${\frak L}$ is well known spherical 
harmonic $Y_L^{(m)}({\bf n})$ with the eigenvalue $\lambda _L$ degenerated with 
respect to the azimuthal number $m=0,\pm 1,...,L$: 
\begin{equation}
{\frak L}Y_L^{(m)}\left( {\bf n}\right) =\int\limits_{\left| {\bf n}%
_1\right| =1}\exp \left[ -\frac{{\cal J}}2|{\bf n}-{\bf n}_1|^2\right]
Y_L^{(m)}\left( {\bf n}_1\right) \frac{d\Omega _1}{4\pi }=\lambda
_LY_L^{(m)}\left( {\bf n}\right) .  
\end{equation}
Applying the addition theorem for spherical harmonics and the eigenfunction
expansion of the solution to Eq.16 one gets closed expressions for the
eigenvalues: 
\begin{equation}
\lambda _L=\Lambda \left(L+\frac{1}{2},{\cal J}\right),\quad \Lambda
\left(\nu,x\right) =\sqrt{\frac {\pi}{2x}}\exp\left( -x\right) 
I_{\nu}\left( x\right) , 
\end{equation}
and the zero moment: 
\begin{equation}
g_{0}({\bf n}_0,{\bf n})=\sum_{L=\,0}^\infty (2L+1)\frac{\lambda _L}{1-\lambda
_L}P_L\left( \cos \theta \right) ,  
\end{equation}
where $I_{\nu}(x)$ is modified Bessel function [14], $P_L(x)$ are the Legendre
polynomials and $\theta $ is the scattering angle: 
$\cos \theta ={\bf n\cdot n}_0$.

Apparent fluctuation intensity depend on the detection direction ${\bf n}$
if the source is {\em directional} (i.e. a narrow range of the radiation
directions ${\bf n}_0$ is allowable). In the opposite case of point source 
all terms except isotropic one with $L=0$ are averaged out when substituting
Eq.19 into Eq.10 at $q=0$. It is the value of ${\cal J}$ that determines how 
strong the anisotropy of Eq.19 is. Power expansion of $I_{\nu}(x)$ with respect
to $x$ leads to $\lambda _L={\cal J}^{\,L}/(2L+1)!!+O\left({\cal J}^{\,L+2}\right)$ 
at ${\cal J}\ll 1$. Thus, at short times Eq.19 is dominated by isotropic term 
$\propto {\cal J}^{\,-2}$, and the anisotropic terms are relatively small as 
${\cal J}^{\,L+2}$. By an analogy with the case ${\cal J}=0, q\neq 0 $ this regime 
may be called {\em diffusional\thinspace} in the sense that multiple scattering retains 
the isotropy of individual scattering on macroscopic scale. Quite opposite behavior 
takes place at long times. Using the asymptotic of $I_{\nu}(x)$ at $x\gg 1$ one obtains 
$\lambda _L=\left(2{\cal J}\right)^{-1}+O\left( {\cal J}^{\,-2}\right)$ at 
${\cal J}\gg 1$. Independence of the leading term in $\lambda _L$ on $L$ makes all 
the terms in the series (19) important. Moreover, since next-to-leading asymptote
of $\lambda _L$ contains the factor $L\left( L+1\right)$ the terms with large orbital
numbers seem to be even more important. This means that the calculation of $g_{0}({\bf n}%
_0,{\bf n})$ requires summation of the whole infinite series. As a result,
scattering angle dependence will be strong.

To exploit classical summation technique the series of interest is
represented as a residue sum over points $Z_L=i\left( L+\frac{1}{2}\right) $,
the zeros of $\cosh \pi z$ 
\begin{equation}
g_{0}({\bf n}_0,{\bf n})=2\pi i\sum_{L=\,0}^{\infty} {\rm Res}\left[ \frac{%
-iz\Lambda \left( -iz,{\cal J}\right) }{1-\Lambda \left( -iz,{\cal J}\right) 
}\cdot \frac{{\rm P}_{-\frac 12-\,iz}\left(\cos \vartheta \right) }{\cosh
\pi z}\right]_{z\,=\,Z_L}\,  
\end{equation}
where $\vartheta =\pi -\theta$, and ${\rm P}_{\nu}\left(x\right)$ is the spherical
function. Adding the sum of residues at $z_l$, the zeros of the function 
$1-\Lambda \left( -iz,{\cal J}\right)$ in upper half-plane, to Eq.20 gives the integral 
of the function in square brackets over the line $\left(-\infty +i\cdot 0,+\infty +i\cdot 0\right)$, 
and therefore 
\[
g_{0}({\bf n}_0,{\bf n})=\int_{-\infty }^{+\infty }\frac{-i\zeta
\Lambda \left( -i\zeta ,{\cal J}\right) }{1-\Lambda \left( -i\zeta ,{\cal J}%
\right) }\cdot \frac{{\rm P}_{-\frac 12-\,i\zeta }\left( \cos \vartheta
\right) }{\cosh \pi \zeta }d\zeta +2\pi \sum_l\frac{z_l\,{\rm P}_{-\frac
12-\,iz_l}\left( \cos \vartheta \right) }{\cosh \left( \pi z_l\right) \frac
d{dz}\Lambda \left( -iz_l,{\cal J}\right) }\cdot 
\]
Here $\Lambda \left( -i\zeta ,{\cal J}\right)$ for real $\zeta$ means
the value on the upper branch of real-axis cut. Because $\overline{\Lambda
\left( -iz\right) }=$ $\Lambda \left( i\,\overline{z}\right) $ both $z_l$
and $-\overline{z_l}$ are the roots. Using this and the properties of the 
{\em cone function\thinspace} P$_{-\frac 12-\,iz}\left( x\right)$ [15] one obtains
after a transformation 
\[
g_{0}({\bf n}_0,{\bf n})=2\int_0^{+\infty }\frac{\zeta {\rm Im}%
\Lambda \left( -i\zeta ,{\cal J}\right) }{\left[ 1-{\rm Re}\Lambda \left(
-i\zeta ,{\cal J}\right) \right] ^2+\left[ {\rm Im}\Lambda \left( -i\zeta ,%
{\cal J}\right) \right] ^2}\cdot \frac{{\rm P}_{-\frac 12-\,i\zeta }\left(
\cos \vartheta \right) }{\cosh \pi \zeta }d\zeta 
\]
\begin{equation}
+4\pi {\rm Re}\sum_{{\rm Re}z_l\,>\,0}\frac{z_l\, {\rm P}_{-\frac
12-\,iz_l}\left( \cos \vartheta \right) }{\cosh \left( \pi z_l\right)
\,\frac d{dz}\Lambda \left( -iz_l,{\cal J}\right) }.  
\end{equation}
With this exact representation the desired asymptotic can be now handled.
Eq.18 and Bessel functions theory [14] give at ${\rm Im}z>0$
\[
\Lambda \left( -iz,{\cal J}\right) =\frac 1{\sqrt{2\pi {\cal J}}%
}\int\limits_0^\pi \,\exp \left[ -{\cal J}\left( 1-\cos u\right) \right]
\cosh \left( uz\right) du
\]
\begin{equation}
+i\frac{\sinh \left( \pi z\right) }{\sqrt{2\pi {\cal J}}}\int\limits_0^%
\infty \exp \left[ -{\cal J}\left( 1+\cosh w\right) +iwz\right] dw.  
\end{equation}
Both integrals in Eq.22 for real $\zeta $ can be estimated at 
${\cal J}\rightarrow \infty $ using the Laplace method. At $\left| \zeta \right| \gg 
\sqrt{2{\cal J}}$ one obtains 
\begin{equation}
\Lambda \left( -i\zeta ,{\cal J}\right) \simeq \frac 1{2{\cal J}}\left[ \exp
\left( \frac{\zeta ^2}{2{\cal J}}\right) +i\sinh \left( \pi \zeta \right)
\exp \left( -2{\cal J-}\frac{\zeta ^2}{2{\cal J}}\right) \right] .  
\end{equation}
Let us assume, in addition, that $\left| \zeta \right| \ll 2 {\cal J}/\pi$. 
The integral in Eq.21 is now easily calculated because under this condition 
lorentzian-like factor in the integrand has exponentially small broadening 
${\rm Im}\Lambda \left(-i\zeta ,{\cal J}\right)$, and hence is 
well approximated by the delta-function 
\[
\frac{\pi \delta \left( \zeta -\zeta _0\right) }{\left| \frac d{dz}{\rm Re}
\Lambda \left( -i\zeta _0,{\cal J}\right) \right| },
\]
where $\zeta _0$ is the positive root of real equation ${\rm Re}\Lambda
\left( -i\zeta ,{\cal J}\right) =1$. Substituting here the real part of 
Eq.23 gives the asymptotic of the root 
\begin{equation}
\zeta _0\simeq \sqrt{2{\cal J}\,\log \left( 2{\cal J}\right) }.
\end{equation}
This equation is consistent with the assumed inequalities provided that 
\begin{equation}
\frac{\sqrt{2{\cal J}}}\pi \gg \sqrt{\log \left( 2{\cal J}\right) }\gg 1. 
\end{equation}
An analysis based Eq.22 shows that in this limit the equation $\Lambda
\left( -iz,{\cal J}\right) =1$ has in the upper right quadrant only one root 
$z_{0}$ given asymptotically by 
\begin{equation}
z_{0}\simeq \zeta _{0}+i\frac{\exp \left( \pi \zeta _{0}-2{\cal J}\right) }
{8\zeta_0{\cal J}}.  
\end{equation}
Due to the inequalities (25) the imaginary part of this root is exponentially
small, and can be discarded when calculating the residue term in Eq.21.
Then the integral and residue contributions give together
\begin{equation}
g_{0}({\bf n}_0,{\bf n})\simeq 6\pi {\cal J}
\frac{{\rm P}_{-\frac{1}{2}-\,i\zeta_0}\left(\cos\vartheta\right)}{\cosh\pi\zeta _0}.
\end{equation}
At the end, using in Eq.27 the asymptotic of cone function at large
argument [15], one obtains for hemisphere$\;0<\theta \leq \pi $ 
\begin{equation}
g_{0}({\bf n}_0,{\bf n})\simeq 12\pi {\cal J}f\left( \theta \right) 
\exp\left[ -\theta \sqrt{2{\cal J}\,\log \left( 2{\cal J}\right) }\right] , 
\end{equation}
where $f\left( \theta \right) =1$ at backward scattering ($\theta =\pi $),
otherwise at 
$2\pi \sin \theta \gg \left[ 2{\cal J}\,\log \left( 2{\cal J}\right) \right] ^{-1/2}$ 
\begin{equation}
f\left( \theta \right) \simeq \left( 2\pi \sin \theta \right) ^{-1/2}
\left[2{\cal J}\,\log \left( 2{\cal J}\right) \right] ^{-1/4}.  
\end{equation}
At forward scattering ($\theta =0$) the whole derivation is inapplicable
because all terms of the series in Eq.19 are positive. In this case the
fluctuation intensity proves to be greater than unity at any time.

To express Eqs.28, 29 via $t$ one may use the models of the scatterer
motion described in the literature. In two simplest, but practically
important, cases of ballistic and diffusive motion: 
$2{\cal J}=\left(t/\tau _{a}\right)^a$, where $a=2$ and $a=1$, respectively.
The formulas for time constants $\tau _a$ are found, for example, in [9,10]. 
Thus, at $t\rightarrow \infty $ the fluctuations decay according to 
power-logarithmic stretched exponential law
\begin{equation}
g_{0}({\bf n}_0,{\bf n})\sim \exp \left[ -\theta \sqrt{a\left(\frac {t}{\tau
_a}\right) ^a\log \left( \frac {t}{\tau _a}\right) }\right] . 
\end{equation} 
\begin{center}
{\large {\bf 5.2. The First and Second Spatial Moments}}
\end{center}

Calculating the asymptotics for $g_\alpha \left( {\bf n}_0,{\bf n}\right) $
and $g_{\alpha \beta }\left( {\bf n}_0,{\bf n}\right) $ is very involved
task. To make things easier one might introduce local coordinate system with 
$z$ - axis parallel to vector ${\bf n}_0$ and consider local longitudinal
components $g_z\left( {\bf n}_0,{\bf n}\right) $ and $g_{zz}\left( {\bf n}_0,%
{\bf n}\right) .$ Other components are zero if spatial variation occurs
only in the radiation direction. This is the case for an infinite source
concentrated at the plane $z=0$. CF calculated using Eq.3 is one half of
the sum of the two attenuated forward and backward plane waves 
\begin{equation}
E_c^{\pm }\left( {\bf r}\right) \propto \exp \left( \pm ik_0z-\frac{\left|
z\right| }{2l}\right) ,  
\end{equation}
that is there are two allowed radiation directions ${\bf n}_0={\bf n}_0^{\pm
}=\left( 0,0,\pm 1\right) $. Fortunately, the superposition principle holds
also for FCF (Eq.10). This turns out to be equal to the sum of half-weighted 
$\gamma \left( {\bf r},t;{\bf n}_0^{\pm },{\bf n}\right) $, both being
independent on $x,y$. Therefore, considering CF and FCF for one direction,
say ${\bf n}_0^{+}$, one could hope to emulate a transmission problem where
plane wave impinges on the boundary $z=0$. Both CF and FCF vanish in the
counterpart half-space, and coherent portion of radiation propagates only
forward, whereas the fluctuations develop in all directions with the intensity 
at $t\rightarrow \infty $ given by Eqs.28, 29.

The first longitudinal moment, calculated recursively as outlined above, is
given by 
\begin{equation}
g_{z}\left( {\bf n}_0,{\bf n}\right) =l\,g_{0}\left( {\bf n}_0,{\bf n}\right)
+l\sum_{L=0}^{\infty}\left[ \frac{\left( L+1\right) \lambda _{L+1}}
{1-\lambda_{L+1}}+\frac{L\lambda _{L-1}}{1-\lambda _{L-1}}\right] 
\frac{P_L\left(\cos\theta \right) }{1-\lambda _L}.  
\end{equation}
At $\theta >0$ and ${\cal J}\rightarrow \infty $ the summation technique
developed above can be applied to Eq.32 as well. After division by Eq.27
one obtains the asymptotic for normalized first moment 
\begin{equation}
\overline{z\,}\simeq \frac{\xi \,\sin \theta }{6},\quad \frac{\xi}{l}=
\sqrt{\frac{2{\cal J}}{\log \left( 2{\cal J}\right) }}=\frac{\left( t/\tau
_a\right) ^{a/2}}{\left[ a\log \left( t/\tau _a\right) \right] ^{1/2}} 
\end{equation}
tending to infinity if $0<\theta <\pi $, and $\overline{z\,}\rightarrow 0$
if $\theta =\pi $. Closed formula for the second longitudinal moment 
$ g_{zz}\left( {\bf n}_0,{\bf n}\right) $ is derived along the same lines 
as Eq.32,and long-time asymptotic of normalized second moment is obtained 
making use the summation method developed above. Expression for 
$g_{zz}\left( {\bf n}_0,{\bf n}\right)$ is rather lengthy, and obtaining 
the asymptotic involves processing many cumbersome terms without bringing 
new ideas. This analysis is omitted, and the final result is as follows 
\begin{equation}
\overline{z\,^2}\simeq \frac{\xi ^2}3=\frac{l^2}{3}\cdot 
\frac{\left( t/\tau_a\right) ^a}{a\log \left( t/\tau _{a}\right) }.  
\end{equation}

In conclusion, whereas the fluctuations intensity decays at $t\rightarrow \infty$ 
according to Eq.30, spatial center of the FCF 'runs away' according to Eq.33, 
and due to Eqs.33, 34 spatial dispersion of the fluctuations 
$\delta =\sqrt{\overline{z\,^2}-\left( \overline{z\,}\right)^2}\simeq %
\xi\left( 1-\sin^{2}\theta/12\right)^{1/2}/\sqrt{3}$ also tends to infinity 
but with much weaker anisotropy. In other words, at long times random motion 
of the scatterers weakens the fluctuations, thus tending the radiation to a 
coherent state, but enhances drastically the spatial range of the field 
correlations $\xi $  over its value for static disorder $\sim l$.

\begin{center}
{\Large {\bf Acknowledgements.}}
\end{center}

I am grateful to Dr. E. Kogan, who attracted my attention to this problem, for
enlightening discussions and critical reading of the manuscript.

\newpage
\begin{center}{\Large {\bf References}}\end{center}

\noindent [1] M.P. van Albada and A. Lagendijk, Phys.Rev.Lett., {\bf 55},
2692 (1985)

\noindent [2] E. Wolf and P. Maret, Phys.Rev.Lett., {\bf 55}, 2696 (1985)

\noindent [3] A.Z. Genack, N. Garcia and W. Polkosnik, Phys.Rev.Lett., {\bf %
65}, 212 (1990)

\noindent [4] M.P. van Albada, J.F. de Boer and A. Lagendijk,
Phys.Rev.Lett., {\bf 64}, 2787 (1990)

\noindent [5] M. Kaveh, Waves in Random Media, {\bf 1}, S121 (1991)

\noindent [6] B. Shapiro, Phys.Rev.Lett. {\bf 57}, 2168 (1986)

\noindent [7] M. Stephen and G. Cwilich, Phys. Rev. {\bf B 34}, 7564 (1986)

\noindent [8] A. Ishimaru,{\em \ Wave Propagation and Scattering in Random
Media}, Vol.1-2 \\(Academic, New York 1978)

\noindent [9] A.A. Golubentsev, Sov. Phys. JETP, {\bf 59}, 26 (1984)

\noindent [10] M. Stephen, Phys. Rev. {\bf B 37}, 1 (1988)

\noindent [11] I. Edrei and M. Kaveh, J. Phys. {\bf C} {\bf 21}, L971 (1988)

\noindent [12] M. Rosenblue, N. Hoshen, I. Freund and M. Kaveh,
Phys.Rev.Lett., {\bf 58}, 2754 (1987)

\noindent [13] E. Kogan, private communication; \\M. Auslender,
Bull. Israel Phys.Soc. {\bf 39}, 134 (1993)

\noindent [14] G.N. Watson, {\em A Treatise on the Theory of Bessel Functions%
},\\(University Press, Cambridge 1966)

\noindent [15] V.V. Lebedev, {\em Special Functions and Their Applications}, %
\\Selected Russian Publications in the Mathematical Sciences, ed. by R.A.
Silverman (Prentice-Hall Inc, Englewood Cliffs, N.J. 1965).
\end{document}